\documentclass[prb,aps,twocolumn,showpacs]{revtex4-1}
\RequirePackage{amsmath}
\usepackage{graphics}
\usepackage{graphicx}
\usepackage{amssymb,amsmath}
\usepackage{bm}

\begin{document}

\title{Non-centrosymmetric superconductors on honeycomb lattice}

\author{ Der-Hau Lee$^{1}$ and Chung-Hou Chung$^{1,2}$}
\affiliation{ $^{1}$Electrophysics Department, National Chiao-Tung
University,
HsinChu, Taiwan, 300, R.O.C. \\
$^{2}$Physics Division, National Center for Theoretical Sciences, HsinChu, Taiwan, 300 R.O.C.\\
 }
\date{\today}

\begin{abstract}
  We study non-centrosymmetric topological superconductivity
in correlated doped quantum spin-Hall insulators (QSHI) on honeycomb
lattice without inversion symmetry where the intrinsic (Kane-Mele)
and Rashba spin-orbit couplings can in general exist.
We explore the generic topologically non-trivial superconducting phase diagram of the model
system.  Over a certain parameter space, the parity-mixing
superconducting state with co-existing spin-singlet $ d$+$id$ and spin-triplet
$p$+$ip$-wave pairing is found. On a zigzag nanoribbon, the parity-mixing superconducting
state shows co-existing helical and chiral Majorana fermions at edges. Relevance of our results
for experiments is discussed.
\end{abstract}


\maketitle   

\section{Introduction}
Non-centrosymmetric supercondutors (NCSs) have drawn intensive
attraction since their discoveries in last decade due to the lack of
space inversion symmetry, such as \begin{math}CePt_3 Si\end{math},
\begin{math}UIr\end{math} and \begin{math}Li_2 Pd_3 B\end{math} \cite{Saxena}.
Based on Pauli principle, the total electronic wave function is
antisymmetric under particle exchange. However, there is no
definitely symmetry of spatial wave function in NCS, giving rise to
the exotic mixture of spin singlet (with even parity) and triplet
states (with odd parity) Cooper pairs. Graphene and graphene-based
two-dimensional materials, such as transition metal dichalcogenides
(TMDs), silicene, germanene, the binary compounds of the group-IV
elements, and the group III.V compounds all have broken inversion
symmetry \cite{MingshengXu}, and are possible candidates of NCS. In
particular, TMDs have been studied extensively both theoretically
and experimentally for its promising properties in valleytronics and
spintronics applications
\cite{TMD1,TMD2,TMD3,TMD4,TMD5,TMD6,TMD7,TMD8,TMD9,TMD10}, such as:
stronger spin-orbit (SO) coupling via $d$ orbital of the metal atoms
in $MoS_2$, strong and robust SO interaction in graphene on
TMD substrates \cite{PRX_16}.\\
\indent For moving electrons in a closed orbit, an electric field
will produce an effective magnetic field via relativistic effect,
which generates a Zeeman energy term,  recognized as Rashba SO
coupling. The electric field based Rashba interaction can be
internal in quantum wells with structural inversion symmetry
breaking or external as is in graphene \cite{Manchon,Ezawa}. Due to
Rashba interaction, these systems show non-trivial helical spin
textures where spins with the same species rotating with the same
chirality at the Fermi surface. Similar spin-momentum-lock phenomena
also occur in topological insulators and NCSs
\cite{Saxena,Manoharan}. Therefore, it is promising to expect the
Rashba coupling to be present in NCS. Note that the geometric Berry
phase of moving electrons in toplological material is associated
with the non-trivial Chern number via Kubo formula for Hall
conductivity \cite{NiuRMP}, indicating the existence of
topologically non-trivial properties, such as: gapless metallic edge
states in topological insulators (TIs) and Majorana fermions (MFs)
as zero-energy charge-neutral edge states of topological
superconductors (TSCs) \cite{Kane10,Zhang11,Jason12,Assaad13}.\\
\indent An interesting aspect of Rashba coupling in NCSs is that it
induces spin-triplet pairing (apart from the conventional spin
singlet pairing) due to the in-balance in populations of different
spin species via Rashba coupling. A non-trivial Chern number has
been predicted in spin triplet $p$+$ip$ superconductors induced by
Rashba interaction on square lattices under strong magnetic fields,
which suggests that NCSs are topological superconductors
\cite{Sato09}. Similar results have been reported on honeycomb
lattice for $MoS_2$, which was predicted to support exotic triplet
pairing phases \cite{KTLaw} and was argued that superconductivity
can be induced there experimentally via applying gate voltages
\cite{Saito_16}.\\
\indent On the other hand, for doped graphene in the absence of SO
couplings, spin-singlet chiral superconducting state with $d$+$id$
pairing is energetically favourable as the ground state
\cite{BlackSchaffer14}. It was found that the $d$+$id$-wave state is
a state with mixed $s$-wave and exotic $p$+$ip$-wave pairing orders
at low energy \cite{JPHu}. The chiral nature comes from fact that
the two $d$+$id$-wave state carry an equal weight under six fold
symmetry of the honeycomb lattice, and a linear combination of two
order parameters is needed to describe the system. The chiral
superconductivity also leads to breaking of both time-reversal (TR)
and parity symmetries. Further investigations confirmed that there
are two co-propagating chiral surface states per zigzag graphene
nanoribbon (ZGNR) edge, which shows the system is topological
non-trivial with Chern number being $2$
\cite{BlackSchaffer}.\\
\indent Including the intrinsic (Kane-Mele) SO coupling in graphene,
exotic and distinct helical topological edge states with two
counter-propagating chiral modes protected by the TR symmetry were
predicted within the Kane-Mele (KM) model, signature of 2D quantum
spin Hall (topological) insulator \cite{Kane05}. The TR symmetry
invariant KM model exhibits a mirror symmetric SO interaction, in
contrast to the Rashba SO coupling \cite{Kane10}, and is a perfect
theoretical model for TI phases in two dimensions \cite{Karyn10}.
Recent study further showed an exotic 2D spin-singlet TSC with
non-trivial pseudospin Chern number in doped correlated KM model on
honeycomb lattice \cite{Sun}. There, the system undergoes a
topological phase transition from a phase with chiral MFs to a phase
with helical Majorana zero modes were realized with increasing the
intrinsic SO coupling.\\
\indent The purpose of this paper is to explore all possible
topologically non-trivial non-centrosymmetric superconducting states
in graphene-based materials by combining all the phenomena mentioned
above, including the effects of intrinsic (KM) and Rashba SO
couplings, the mixture of $d$+$id$ spin singlet and triplet
$p$+$ip$-wave pairings in the presence of both  KM and Rashba SO
couplings. A recent related study on the simplified model without KM
interaction in cuprates has shown non-trivial topological properties
\cite{Yoshida}. However, there is still lack of a comprehensive
analysis of the more general and generic model systems for NCS on
honeycomb lattice. To realize parity mixing phenomena, experimental
setups have been proposed theoretically \cite{Yoshida,exp1,exp2}.
The corresponding Majorana mode can be detected by Andreev
reflection \cite{AR} or electrical detection \cite{AR2} where some
signatures of MFs have been observed experimentally in hybrid
structures via these approaches \cite{exp3,exp4}. The remaining
parts of the paper are organized as follows: Section 2 describes
computation model and Chern number calculation formula. Section 3
presents main results: the topological phase diagram and
corresponding edge spectrum. In this work, we aim for edge states of
graphene-based system, which is zigzag nanoribbon (ZNR). Section 4
summarizes results and discusses related issues. Finally,
conclusions are provided in Section 5.

\begin{figure}[t]%
\includegraphics*[width=0.475\textwidth]{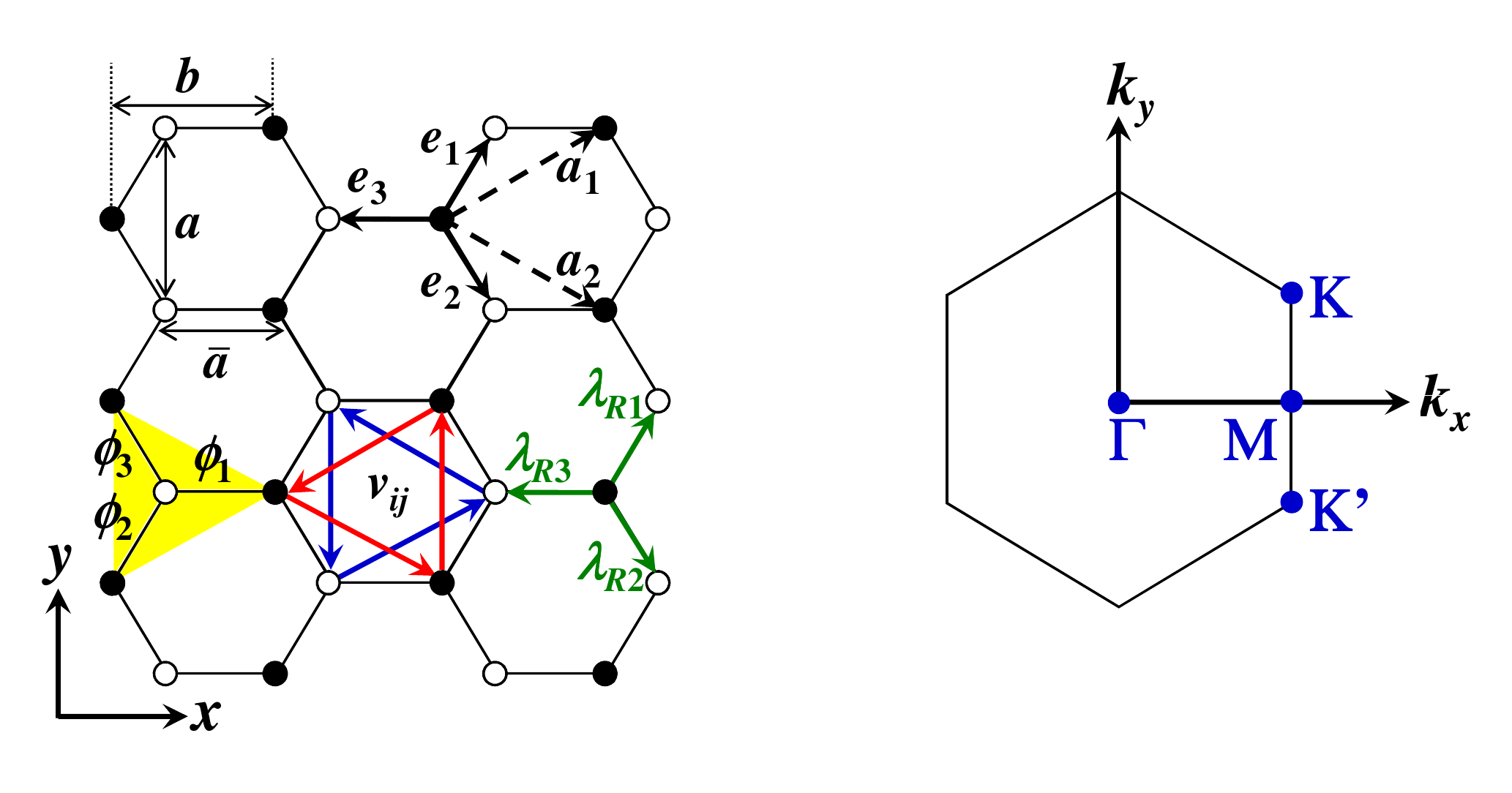}
\caption{(Left) Plot for honeycomb lattice. (right) Corresponding first Brillouin
zone with the two inequivalent Dirac points $K$ and $K'$ and high symmetry points $\Gamma$  and $M$. The honeycomb lattice consists of two interpenetrating
triangular lattices denoted by sublattice A (dark circles) and sublattice
B (open circles) with lattice vectors $a_1$ and $a_2$ (dashed arrows).
The nearest-neighbor lattice vectors between nearest-neighbor A and
B sites are denoted by $e_{i=1,2,3}$ with site-site distance $\bar a$ and lattice constant $a = \sqrt 3 \bar a$.
Here, $\bar a$ is set to be 1. $b$ is used
to label the location of individual zigzag chain along $x$ axis. The red
(blue) arrows within sublattice A(B) represent the directions of the
next-nearest-neighbor hopping term of Kane-Mele model. The green
arrows indicate the spin dependent interaction of Rashba model, and shaded
yellow triangle represents amplitudes of the superconducting pairing gaps (see text).}
\label{fig1}
\end{figure}

\section{The model}

The Rashba and KM SO couplings have been argued to exist in graphene
\cite{Kane05}. Both Rashba and KM interactions couple spin-up and
spin-down states, and break the SU(2) symmetry. However, the U(1)
symmetry still remains for the former, whereas the latter does not
\cite{StephanRachel}.

\subsection{Rashba spin-orbital interaction}
For electric field perpendicular to the graphene plane, the
Hamiltonian of Rashba model is represented as \cite{NancySandler}

\begin{equation}
H_R  = i\lambda _R\sum\limits_{\left\langle {ij} \right\rangle {\sigma} {\sigma '} } { a_{i{\sigma } }^{\dagger} \left( {{\tilde \sigma } \times {  e }_{ij} } \right)_z b_{j{\sigma '} }^{} }  + h.c..
\end{equation}
Here, $a(b)$ stand for sublattices A(B), $a_{i\sigma }^\dagger
\left( {b_{j{\sigma '} } } \right) $ is a creation (an annihilation)
operator for an electron at site $i$($j$) with spin $\sigma$
(${\sigma '}$). The Pauli matrix is defined as $\tilde \sigma  =
\left( {\sigma _x ,\sigma _y ,\sigma _z } \right) $. $\lambda _R$ is
the Rashba coupling. The nearest-neighbor (NN) vectors of the
honeycomb lattice are given by: $(1/2,\sqrt 3 /2)$,
$(1/2,-\sqrt 3/2)$, and $(-1,0)$.
$ \left\langle {ij}\right\rangle$ denotes NN sites, connected
by unit vectors ${{e}_{ij} }$. ${\lambda _R }$ is proportional to the electric field.
The explicit Rashba SO terms, ${\left( {{\tilde \sigma } \times {
e}_{ij} } \right)_z }$, along three NN vectors, ${{ e}_i }$, can be
described by

\begin{equation}
\begin{array}{l}
 \lambda _{R1}  = {{\left( {\sigma _y  - \sqrt 3 \sigma _x } \right)} \mathord{\left/
 {\vphantom {{\left( {\sigma _y  - \sqrt 3 \sigma _x } \right)} 2}} \right.
 \kern-\nulldelimiterspace} 2}, \\
 \lambda _{R2}  = {{\left( {\sigma _y  + \sqrt 3 \sigma _x } \right)} \mathord{\left/
 {\vphantom {{\left( {\sigma _y  + \sqrt 3 \sigma _x } \right)} 2}} \right.
 \kern-\nulldelimiterspace} 2}, \\
 \lambda _{R3}  =  - \sigma _y,  \\
 \end{array}
\end{equation}
those are illustrated schematically as green lines in figure 1.

The bulk band structure of graphene under Rashba interaction does
not open a gap in the spectrum. It exhibits the crossings of
conduction and valence bands, giving rise to the splitting of the
original dispersion line on the Dirac points. The probability of
finding an electron in the spin-up and spin-down state is equal. As
a result, the Rashba SO interaction does not break TR symmetry. For
the band structure of ZGNR, it has be shown that there is crossing
of conduction and valence edge bands, whereas intrinsic SO coupling
tend to open gaps. Furthermore, the Rashba SO interaction produces a
non-homogeneous spin polarization on the edge states of the ZGNR.
This is in contrast to the effect produced by the intrinsic KM SO
coupling where spin-polarized states have same spatial spin
distribution \cite{NancySandler}.

\subsection{Kane-Mele (intrinsic) spin-orbital interaction}

The Kane-Mele SO coupling is represented by the spin dependent
next-nearest-neighbor (NNN) hopping, reads \cite{Kane05}

\begin{equation}
H_{KM}  = i\lambda _{SO} \sum\limits_{\left\langle {\left\langle {ij} \right\rangle } \right\rangle \sigma } {v_{ij} {{\sigma _z } c_{i\sigma }^{\dagger} c_{j\sigma} }  + h.c..}
\end{equation} Here, $c = a$ or $b$. $\lambda _{SO}$ is the Kane-Mele (intrinsic)
spin-orbital coupling. $\left\langle {\left\langle {ij}
\right\rangle } \right\rangle $ denotes NNN sites. For the same
sublattice hopping, amplitude ${v_{ij} }$ = 1 or $-$1 (red
counter-clockwise or blue clockwise arrows in figure 1)
\cite{PRB2014}.

The Haldane model of honeycomb lattice shows that a non-trivial
topology can exist when TR symmetry is broken. The KM model can be
considered as spinful Haldane model, composed of two decoupled
Hamiltonians, there up and down spin electrons exhibit helical
quantum Hall effects in graphene, and backscattering are forbidden
due to TR invariance \cite{Hall}. The band structure of a ZGNR with
KM interaction shows that there are two bands which traverse the
bulk gap, connecting the $K$ and $K'$ points. These bands are
localized edge states. The exotic edge states are helical, which
propagate in both directions in each band. There, electrons with
opposite spin propagate in opposite directions, leading to the spin
filtered states \cite{Kane05}. However, a pair of counterpropagating
edge modes at each edge can acquire a gap in the presence of large
$\lambda _{SO}$ in KM model \cite{Karyn10}. As a result, this system
needs further $Z_2$ topological classification. It is shown that
system with KM and Rashba interaction in the presence of magnetic
fields shows the QAH phase with Chern number $C= 2 $ \cite{Niu}.
Furthermore, the above model was shown to exhibited equivalent
mathematical structure to a spin singlet $d$+$id$ superconductor via
a duality mapping \cite{PRB2016}.

\subsection{Possible superconducting states in graphene-based materials}
For graphene without SO coupling, it has been shown that the
superconducting state is of $d$+$id$ spin singlet pairing at low
energy. By examining the superconducting order on $K$ and $K'$
points, $s$-wave and exotic $p$+$ip$-wave pairing orders emerge,
indicating that $d$+$id$-wave pairing is a mixed state from these
two orders \cite{JPHu}. For graphene with KM interaction in stripe
geometry, the system is found to also exhibit chiral singlet $d$+$id$-wave
superconductivity near half-filling under TR symmetry breaking via a
renormalized mean-field approach to the $t-J$ model. For strong KM interaction,
despite the TR symmetry breaking $d+id'$ superconducting state, helical MFs in 2D spin-singlet
topological superconducting state are still found,
protected by a pseudo-spin symmetry.
With decreasing KM interaction, the system
undergoes a topological phase transition to a phase with chiral MFs \cite{Sun}. To go a
step further, our focus here is to study possible superconducting
states when both KM and Rashba terms are present. We search for
possible co-existing phase between singlet $d$+$id$- and triplet
$p$+$ip$-wave superconductivity.\\
\indent We now discuss mechanism of possible triplet $p$+$ip$-wave
superconductivity on honeycomb lattice. The superconducting pairing
on honeycomb lattice has been proposed to exist between electrons on
NN sites. The three amplitudes of the superconducting pairing gaps
along the three unit lattice vectors are defined as: (1,
$e^{i{2\pi}/3}$, $e^{i{4\pi}/3}$) for $d$+$id$-wave pairing,
$(-2,1,1)$ and $(0,1,-1)$ for $p$+$ip$-wave pairing  (see shaded
yellow triangle in figure 1) \cite{SCG5}, which has already been
revealed in RVB order parameter \cite{BlackSchaffer07}. Notice that
the spin triplet $p$+$ip$-wave amplitude comes from the linear
combination of two components in
irreducible representation $E_2$ \cite{SCG4,SCG7}.\\
\indent The spin triplet $p$+$ip$-wave paring symmetry in honeycomb
lattice is linked to the Rashba coupling. In
Ref.\cite{BlackSchaffer07}, the Hamiltonian for superconducting
graphene, including nearest-neighbor hopping and an
Resonating-Valence-Bond (RVB) interaction term, is treated at a
mean-field level. In general, it was found that the pairing
amplitude along NN direction is $(1,1,1)$ for $s$ wave pairing,
$(0,-1,1)$ and $(2,-1,-1)$ for $d_{xy}$/$p_x$ and $d_{x^2 - y^2
}$/$p_y$ pairing state, respectively. Moreover, the complex linear
combination of $d/p$ vectors leading to the $d$+$id$-/$p$+$ip$-
pairing state \cite{SCG5}. It can be seen that, the component of the
Rashba term in three different hopping directions shown in Eq. (2)
is proportional to that of $p_x$/$p_y$ pairing state with odd parity
symmetry. They are to be combined to form the $p$+$ip$ pairing state
due to the Rashba interaction itself. In addition, previous studies
show that the spin-triplet component is aligned with the Rashba
coupling through the linearized gap equation \cite{Sato09,PRL2004}.
It has also been shown that the triplet pairing is enhanced by
increasing the Rashba coupling \cite{Yu}: the amplitude of triplet
pairing state with $S^z$=$\pm 1$ component dominates in the large
Rashba coupling regime, confirming that the Rashba coupling favours
spin-triplet superconductivity. These developments form a basis for
our further investigation on NCS in graphene-based materials.\\
\indent The general bulk Hamiltonian with $d + p$ mixed pairing
symmetry we consider includes a nearest hopping term ($H_t$), KM
interaction ($H_{KM}$), Rashba interaction ($H_{R}$), the mixing singlet
$d$+$id$-wave and triplet $p$+$ip$-wave pairing ($H_{\Delta}$),
chemical potential term ($H_{\mu}$), and Zeeman coupling term
($H_B$). In momentum space, the model Hamiltonian of the system is
written as $H = H_t  + H_{KM}  + H_R      + H_\mu   + H_B +
H_{\Delta }$, with

\begin{equation}
\begin{array}{l}

H_t  = t\sum\limits_{k\sigma } {g_k a^\dagger_{k\sigma
} b_{k\sigma } }  + h.c.,\\

H_{KM}  = \lambda _{SO} \sum\limits_{k\sigma } {\gamma _k \sigma _z
\left( {a_{k\sigma }^\dagger a_{k\sigma }  - b_{k\sigma }^\dagger
b_{k\sigma } }
\right)} ,\\

H_R  = \lambda _R \sum\limits_{k\sigma\sigma '} \tilde {R_k} \cdot
\tilde \sigma a^\dagger_{k\sigma }
b_{k\sigma '}  + h.c.,\\

H_\mu   = \mu \sum\limits_{k\sigma } \left({a^\dagger_{k\sigma } a_{k\sigma } + b^\dagger_{k\sigma } b_{k\sigma } }\right),\\

H_B  = \mu _B B_z \sum\limits_{k\sigma } { {\sigma _z }
 \left(a^\dagger_{k\sigma } a_{k\sigma '}
 + b^\dagger_{k\sigma } b_{k\sigma '}\right) },\\

 H_{\Delta}
 = \frac{1}{2}\sum\limits_{k\sigma \sigma '} {\left[ {\Delta \left( k \right)a^\dagger_{k\sigma } b^\dagger_{ - k\sigma '}
 + \Delta^* \left( k \right)a_{ - k\sigma } b_{k\sigma '} } \right]}.\\
 \end{array}
\end{equation} Here $k = \left( {k_x ,k_y } \right) $, and $t$, $\mu$ are the NN hopping parameter and chemical potential, respectively.
The magnetic field $B_z$ is in the $z$ direction, which results in
the Zeeman coupling ($\mu _B B_z$). In what follows we set $t = 1$.
The $k$-dependent NN hopping amplitude, KM and Rashba couplings are
given by

\begin{equation}
\begin{array}{l}
 
 g_k  = \sum\limits_{l = 1}^3 {e^{ik \cdot e_l } } ,\\
     \gamma _k  = \sum\limits_{l = 1}^6 {e^{i\left( {k \cdot e'_l  + l\pi  + {\pi  \mathord{\left/
 {\vphantom {\pi  2}} \right.
 \kern-\nulldelimiterspace} 2}} \right)} } ,\\

\tilde R_k  = \sum\limits_{l = 1}^3 {\left( { - e_{l,y} ,e_{l,x} }
\right)e^{i\left( {k \cdot e_l  + {\pi  \mathord{\left/
 {\vphantom {\pi  2}} \right.
 \kern-\nulldelimiterspace} 2}} \right)} },\\
 \end{array}
\end{equation} where $e_l  = \left( {e_{l,x} ,e_{l,y} } \right) $. $e'_{l}$ are
NNN vectors, given by $e'_{1,2}  =  \pm a_1$, $e'_{3,4}  =  \mp
a_2$, and $e'_{5,6}  =  \pm \left( {a_2  - a_1 } \right)$. The gap
function can be represented as $ \Delta \left( k \right) = i\Delta
_{d} \sigma _y + i\Delta _t {\tilde R_k } \cdot {\tilde \sigma }
\sigma _y$ for the general co-existing singlet ($d$+$id$-wave) and
triplet ($p$+$ip$-wave) pairing state, while $\Delta _{d}$ and
$\Delta _{t}$ are the corresponding parameters, respectively. In the
Nambu basis, the Hamiltonian Eq. (4) takes the form $H =
\sum\limits_{ k} {\psi ^ {\dag} {\tilde h\left( k \right)}\psi } $,
where

\begin{equation}
 \psi ^{\dag} = \left(
{\begin{array}{*{20}c}
   {a_{k \uparrow }^{\dag} } & {b_{k \uparrow }^{\dag} } & {a_{k \downarrow }^{\dag} } & {b_{k \downarrow }^{\dag} } & {a_{ - k \uparrow }^{} } & {b_{ - k \uparrow }^{} } & {a_{ - k \downarrow }^{} } & {b_{ - k \downarrow }^{} }  \\
\end{array}} \right).
\end{equation} The 8$\times$8 matrix ${\tilde h\left( k \right)}$ reads

\begin{equation}
\tilde h\left( k \right) = \left( {\begin{array}{*{20}c}
   {\tilde h_{k,k} } & {\tilde h_{-k,  k} }  \\
   {\tilde h_{ k,-k} } & {\tilde h_{ - k, - k} }  \\
\end{array}} \right),
\end{equation} with

\begin{equation}
\tilde h_{k,k}  = \left( {\begin{array}{*{20}c}
   {\gamma '_k  - \mu _ -  } & { - g_k } & 0 & { - \lambda _R R_{ - k} }  \\
   { - g_k^* } & { - \gamma '_k  - \mu _ -  } & {\lambda _R R_k } & 0  \\
   0 & {\lambda _R R_k^* } & { - \gamma '_k  - \mu _ +  } & { - g_k }  \\
   {- \lambda _R R_{ - k}^* } & 0 & { - g_k^* } & {\gamma '_k  - \mu _ +  }  \\
\end{array}} \right),
\end{equation}

\begin{equation}
\tilde h_{ - k, - k}  = \left( {\begin{array}{*{20}c}
   {\gamma '_k  + \mu _ -  } & { - g_k } & 0 & {\lambda _R R_k^* }  \\
   { - g_k^* } & { - \gamma '_k  + \mu _ -  } & { - \lambda _R R_{ - k}^* } & 0  \\
   0 & { - \lambda _R R_{ - k} } & { - \gamma '_k  + \mu _ +  } & { - g_k }  \\
   {\lambda _R R_k } & 0 & { - g_k^* } & {\gamma '_k  + \mu _ +  }  \\
\end{array}} \right),
\end{equation}

\begin{equation}
\tilde h_{-k, k}  = \left( {\begin{array}{*{20}c}
   0 & {\Delta _t R_{ - k} } & 0 & {\Delta _d L_k }  \\
   { - \Delta _t R_k } & 0 & {\Delta _d L_{ - k} } & 0  \\
   0 & { - \Delta _d L_k } & 0 & {\Delta _t R_k^* }  \\
   { - \Delta _d L_{ - k} } & 0 & { - \Delta _t R_{ - k}^* } & 0  \\
\end{array}} \right),
\end{equation}

\begin{equation}
\tilde h_{ k,-k}  = \left( {\begin{array}{*{20}c}
   0 & { - \Delta _t R_k^* } & 0 & { - \Delta _d L_{ - k}^* }  \\
   {\Delta _t R_{ - k}^* } & 0 & { - \Delta _d L_k^* } & 0  \\
   0 & {\Delta _d L_{ - k}^* } & 0 & { - \Delta _t R_{ - k} }  \\
   {\Delta _d L_k^* } & 0 & {\Delta _t R_k } & 0  \\
\end{array}} \right).
\end{equation} Then, the matrix elements of the Hamiltonian can be written as

\begin{equation}
\begin{array}{l}
     L_k  = \sum\limits_{l = 1}^3 {e^{i\left[ {k \cdot e_l  + {{2\pi \left( {l - 1} \right)} \mathord{\left/
 {\vphantom {{2\pi \left( {l - 1} \right)} 3}} \right.
 \kern-\nulldelimiterspace} 3}} \right]}},\\

 R_k = - e^{ik_x }  + e^{{{ - ik_x } \mathord{\left/
 {\vphantom {{ - ik_x } 2}} \right.
 \kern-\nulldelimiterspace} 2}} \left[ {\cos \left( {\frac{{\sqrt 3 k_y }}{2}} \right) - \sqrt 3 \sin \left( {\frac{{\sqrt 3 k_y }}{2}} \right)} \right],\\

   \gamma '_k = \lambda _{SO}\gamma_k,\\


\mu _ \pm   = \mu  \pm \mu _B B_z.\\

 \end{array}
\end{equation} Here $L_k$ and $R_k$ are $k$-components of the Fourier
transformed $d$+$id$-wave and $p$+$ip$-wave pairing gaps,
respectively. $\mu _ \pm$ is the modified chemical potential of spin
up and down electrons. In two dimensional system, inversion and TR
symmetries give rise to gapless Dirac modes since these symmetries
enforce the $\sigma _z $ terms vanish in the Hamiltonian
\cite{Bernevig}. For matrix $\tilde h\left( k \right) $, KM model
shows TR symmetry, whereas $d$+$id$ and $p$+$ip$-wave order
parameters, and Rashba SO interaction break the TR symmetry. In
addition, the Rashba coupling and $p$+$ip$-wave order parameters
break inversion symmetry \cite{Sun,Kao}.\\
\indent To simplify complex bulk Hamiltonian with Rashba
interaction, the author in Ref.\cite{BlackSchaffer} takes symmetric
and antisymmetric combinations of operators $a$($a^\dag$) and
$b$($b^\dag$) as new band operators. The resulting Hamiltonian is
composed of lower and upper $\pi$ band. The lower band Bogoliubov-de
Gennes (BdG) Hamiltonian can be diagonalized through standard
Bogoliubov transformation to determine the electronic structure of
Fermi surface.\\

\subsection{Chern number calculation}
 To search for edge modes, the Chern number (Hall conductance) of the system with periodic
  boundary conditions (bulk) is considered. The Chern number is obtained by integrating Berry curvature over
all occupied bands in first Brillouin zone (BZ), can be expressed in
the form \cite{Vafek,Thouless}

\begin{align}
\begin{split}
    C ={}&  \frac{{ - 1}}{\pi }\int\int {dk_x dk_y }\\
         &  \sum\limits_{E_\alpha   < 0 < E_\beta  } {\frac{{{\mathop{\rm  Im}\nolimits} \left\langle \alpha  \right|\partial _{k_x } \tilde h\left( k \right)\left| \beta  \right\rangle \left\langle \beta  \right|\partial _{k_y } \tilde h\left( k \right)\left| \alpha  \right\rangle }}{{\left( {E_\alpha   - E_\beta  } \right)^2 }}},
\end{split}
\end{align} where $\alpha$ and $\beta$ denote the quasiparticle bands, and $C$
is the charge Chern number. Note that in the presence of a finite
magnetic field and Rashba coupling, $S^z$ is not a good quantum
number \cite{xu}. Thus, above equation is not suitable for describing the spin Chern number. Other approaches are needed, which go beyond the scope of this paper.\\

\begin{figure}[t]%
\includegraphics*[width=0.475\textwidth]{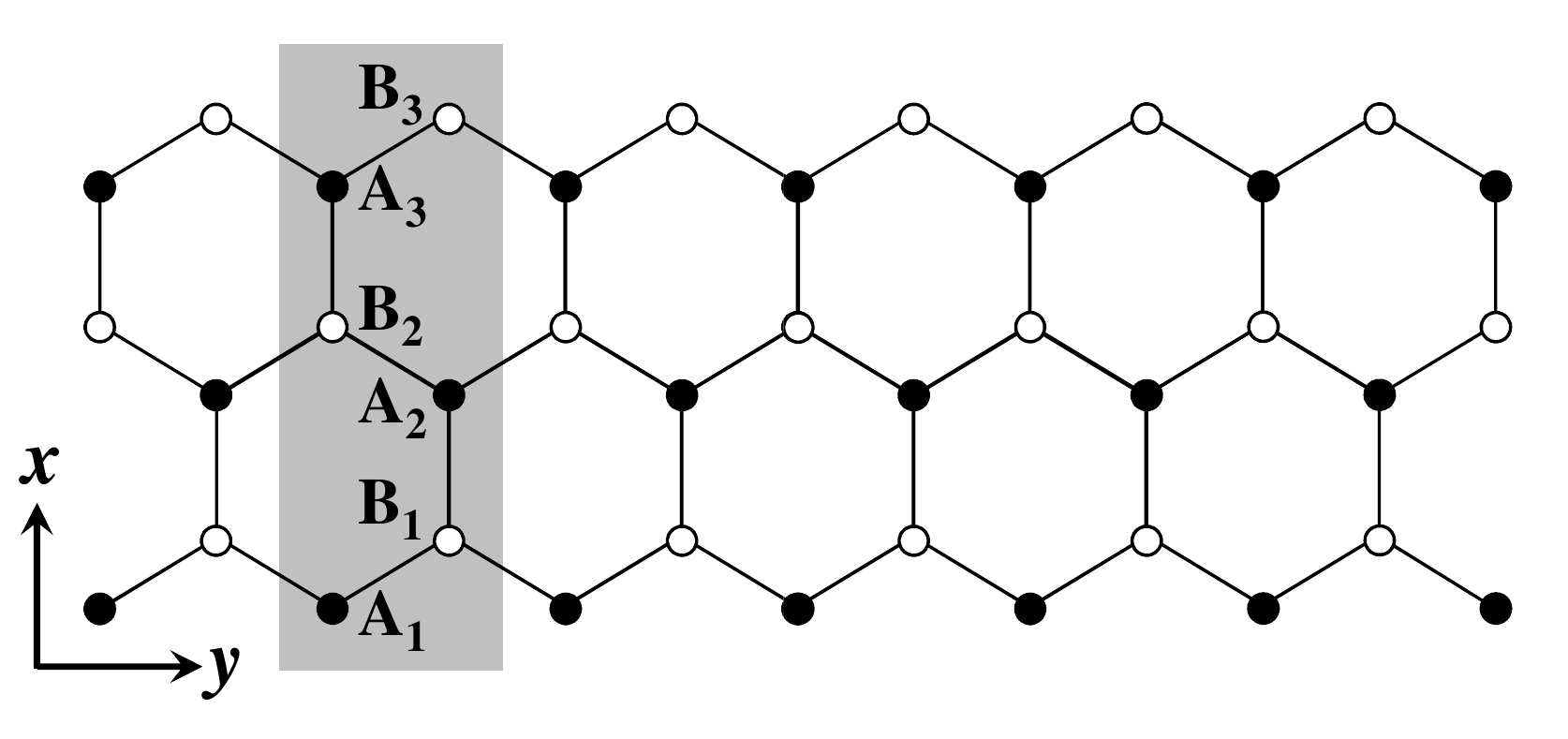}
\caption{Schematic plot for a zigzag ribbon. This system shows translational symmetry along $y$ axis. The gray shaded region represents
 for the super unit cell. Here, there are three zigzag chains along
$x$ axis, denoted as 3-ZNR. Each super unit cell is composed of six sites, A$_{1,2,3}$ and B$_{1,2,3}$ for sublattice A and B, respectively.}
\label{fig1}
\end{figure}

\subsection{Hamiltonian matrix of finite system}
In principle, the edge states connect bulk bands inside the band
gap, and the number of edge modes is equal to the sum of the Chern
numbers of each band below Fermi level. To analyze edge modes of the
honeycomb lattice, a ZNR is considered. The Hamiltonian matrix is
diagonalized directly to obtain band structure. For simplicity, the
Hamiltonian matrix for a 3-ZNR within this formalism is given below.
This can be straightforwardly extended to ZNR with different widths.

The quantum many body state can be written in the following basis:

\begin{equation}
\Phi ^\dag  = \left( {\begin{array}{*{20}c}
   {\phi _{k \uparrow }^\dag } & {\phi _{k \downarrow }^\dag } & {\phi _{ - k \uparrow }^{} } & {\phi _{ - k \downarrow }^{} }  \\
\end{array}} \right),
\end{equation} each substate passes through either
A or B sites of all zigzag chains (see figure 2), reads $ \phi  =
\left( {\begin{array}{*{20}c}
   {A_1 } & {B_1 } & {A_2 } & {B_2 } & {A_3 } & {B_3 }  \\
\end{array}} \right)$ \cite{PRB2014}. Hence the Hamiltonian matrix in this basis has the form

\begin{equation}
{\tilde H} = \left( {\begin{array}{*{20}c}
   {{\tilde H}_{k \uparrow ,k \uparrow } } & {{\tilde H}_{k \downarrow ,k \uparrow } } & {{\tilde H}_{ - k \uparrow ,k \uparrow } } & {{\tilde H}_{ - k \downarrow ,k \uparrow } }  \\
   {{\tilde H}_{k \uparrow ,k \downarrow } } & {{\tilde H}_{k \downarrow ,k \downarrow } } & {{\tilde H}_{ - k \uparrow ,k \downarrow } } & {{\tilde H}_{ - k \downarrow ,k \downarrow } }  \\
   {{\tilde H}_{k \uparrow , - k \uparrow } } & {{\tilde H}_{k \downarrow , - k \uparrow } } & {{\tilde H}_{ - k \uparrow , - k \uparrow } } & {{\tilde H}_{ - k \downarrow , - k \uparrow } }  \\
   {{\tilde H}_{k \uparrow , - k \downarrow } } & {{\tilde H}_{k \downarrow , - k \downarrow } } & {{\tilde H}_{ - k \uparrow , - k \downarrow } } & {{\tilde H}_{ - k \downarrow , - k \downarrow } }  \\
\end{array}} \right),
\end{equation} where submatrices take the following form:

\[
\begin{array}{l}
 {\tilde H}_{k \uparrow ,k \uparrow }  = {\tilde H}_t  + {\tilde H}_{KM}  - {\tilde H}_\mu   + {\tilde H}_B  {,}\\
 {\tilde H}_{k \downarrow ,k \downarrow }  = {\tilde H}_t  + {\tilde H}_{KM} \left( {\lambda _{SO}  \to  - \lambda _{SO} } \right) - {\tilde H}_\mu   - {\tilde H}_B  ,\\
 {\tilde H}_{ - k \uparrow , - k \uparrow }  =  - {\tilde H}_t  - {\tilde H}_{KM} \left( {\lambda _{SO}  \to  - \lambda _{SO} } \right) + {\tilde H}_\mu   + {\tilde H}_B  ,\\
 {\tilde H}_{ - k \downarrow , - k \downarrow }  =  - {\tilde H}_t  - {\tilde H}_{KM}  + {\tilde H}_\mu   - {\tilde H}_B  ,\\
 {\tilde H}_{k \downarrow ,k \uparrow }  = {\tilde H}_R  ,\\
 {\tilde H}_{k \uparrow ,k \downarrow }  = {\tilde H}_{k \downarrow ,k \uparrow } \left( {\gamma _1  \to  - \gamma _1 } \right) ,\\
 {\tilde H}_{ - k \downarrow , - k \uparrow }  = {\tilde H}_{k \uparrow ,k \downarrow } , \\
 {\tilde H}_{ - k \uparrow , - k \downarrow }  = {\tilde H}_{k \downarrow ,k \uparrow }  ,\\

 {\tilde H}_{k \downarrow , - k \downarrow }  = {\tilde H}_{\Delta _t }  ,\\
 {\tilde H}_{k \uparrow , - k \uparrow }  =  - {\tilde H}_{k \downarrow , - k \downarrow } \left( {\gamma _1  \to  - \gamma _1 } \right) ,\\
 \tilde H_{ - k \downarrow ,k \downarrow }  = \tilde H_{k \downarrow , - k \downarrow }^{^\dag} , \\
 \tilde H_{ - k \uparrow ,k \uparrow }  = \tilde H_{k \uparrow , - k \uparrow }^{^\dag} , \\
 {\tilde H}_{k \uparrow , - k \downarrow }  = {\tilde H}_{\Delta _d }  ,\\
 {\tilde H}_{k \downarrow , - k \uparrow }  =  - {\tilde H}_{k \uparrow , - k \downarrow }  ,\\
 \tilde H_{ - k \downarrow ,k \uparrow }  = \tilde H_{k \uparrow , - k \downarrow }^{^\dag} , \\
 \tilde H_{ - k \uparrow ,k \downarrow }  = \tilde H_{k \downarrow , - k \uparrow }^{^\dag} , \\
 \end{array}
\] with

\begin{equation}
{\tilde H}_t  = \left( {\begin{array}{*{20}c}
   0 & {t\alpha } & {} & {} & {} & {}  \\
   {t\alpha } & 0 & t & {} & {} & {}  \\
   {} & t & 0 & {t\alpha } & {} & {}  \\
   {} & {} & {t\alpha } & 0 & t & {}  \\
   {} & {} & {} & t & 0 & {t\alpha }  \\
   {} & {} & {} & {} & {t\alpha } & 0  \\
\end{array}} \right),
\end{equation}

\begin{equation}
{\tilde H}_{KM}  = \left( {\begin{array}{*{20}c}
   { - \beta _{\rm 2} } & 0 & { - \beta _{\rm 1} } & {} & {} & {}  \\
   0 & {\beta _{\rm 2} } & 0 & {\beta _{\rm 1} } & {} & {}  \\
   { - \beta _{\rm 1} } & 0 & { - \beta _{\rm 2} } & 0 & { - \beta _{\rm 1} } & {}  \\
   {} & {\beta _{\rm 1} } & 0 & {\beta _{\rm 2} } & 0 & {\beta _{\rm 1} }  \\
   {} & {} & { - \beta _{\rm 1} } & 0 & { - \beta _{\rm 2} } & 0  \\
   {} & {} & {} & {\beta _{\rm 1} } & 0 & {\beta _{\rm 2} }  \\
\end{array}} \right),
\end{equation}

\begin{equation}
{\tilde H}_R  = \left( {\begin{array}{*{20}c}
   0 & {\lambda _R \gamma '} & {} & {} & {} & {}  \\
   {\lambda _R \gamma ''} & 0 & {\lambda _R \gamma } & {} & {} & {}  \\
   {} & { - \lambda _R \gamma } & 0 & {\lambda _R \gamma '} & {} & {}  \\
   {} & {} & {\lambda _R \gamma ''} & 0 & {\lambda _R \gamma } & {}  \\
   {} & {} & {} & { - \lambda _R \gamma } & 0 & {\lambda _R \gamma '}  \\
   {} & {} & {} & {} & {\lambda _R \gamma ''} & 0  \\
\end{array}} \right),
\end{equation}

\begin{figure*}[!t]%
{%
\includegraphics*[width=0.33\linewidth]{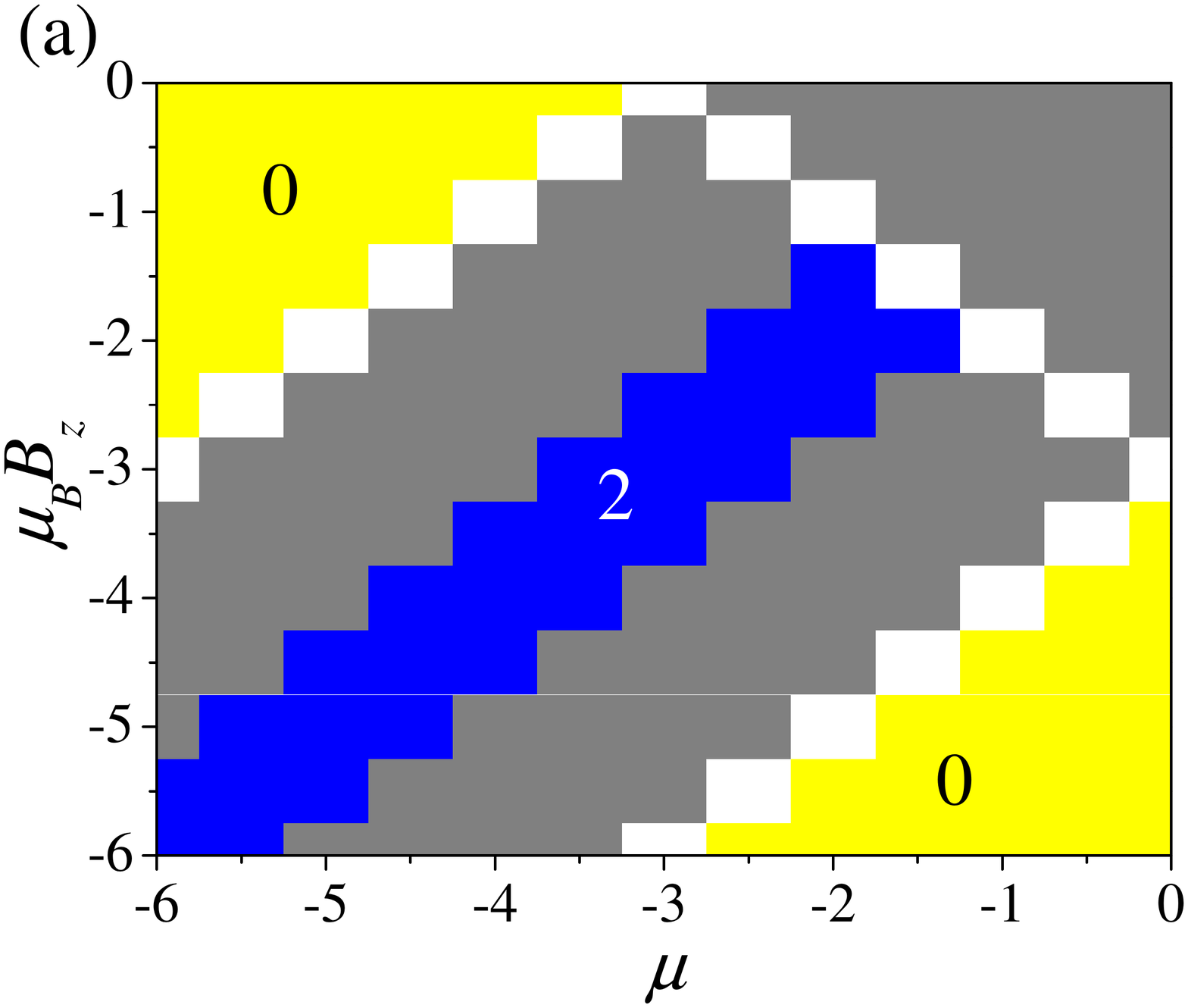}}\hfill
{%
\includegraphics*[width=0.33\linewidth]{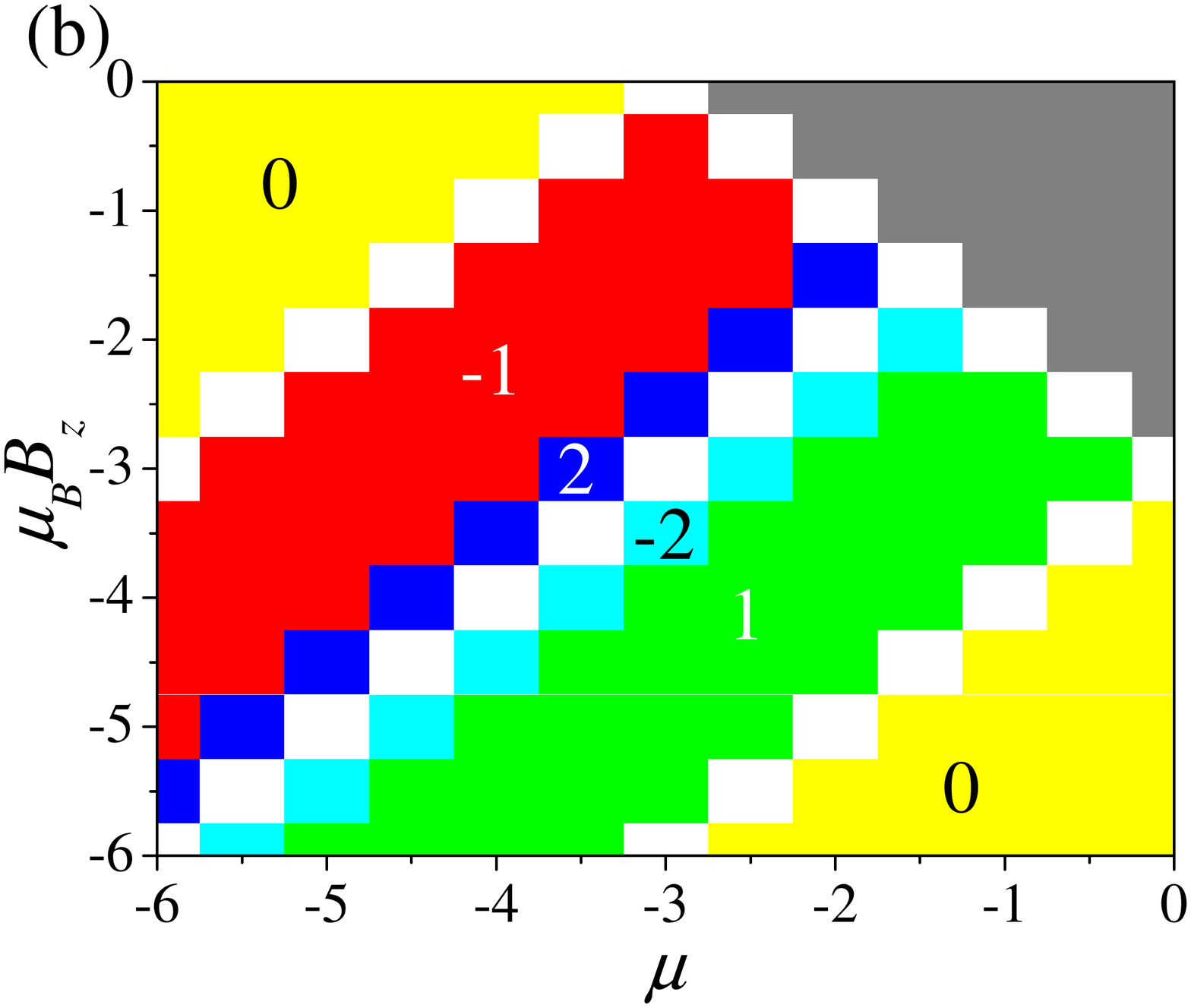}}\hfill
{\label{fig2}%
\includegraphics*[width=0.33\linewidth]{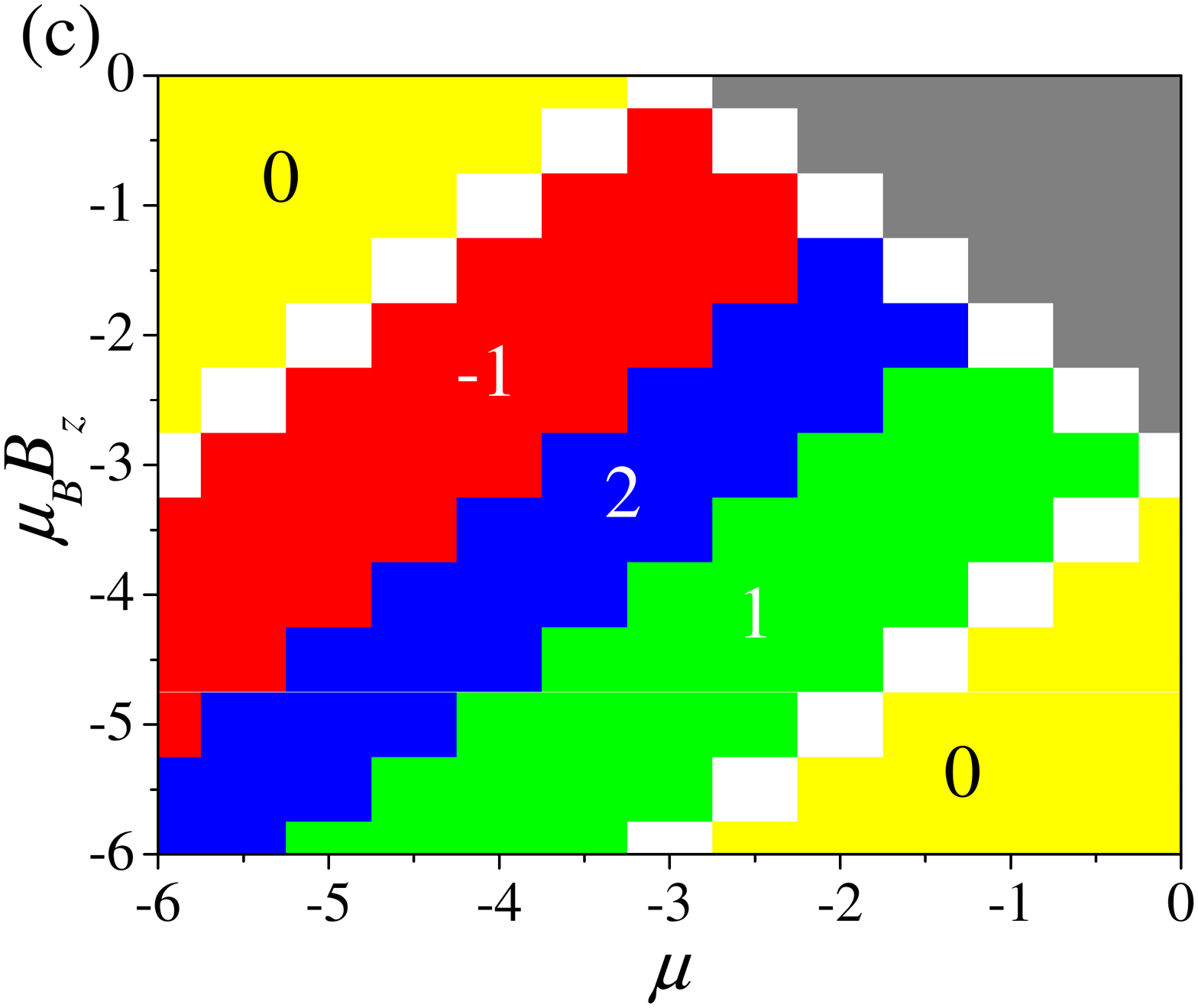}}%
\caption{Topological phase diagram for different pairing sates. (a)
Singlet: $\lambda _R = 0$, $\Delta _t  = 0$, $\Delta _d  = -0.33$,
$\lambda _{SO}  = 0.143$. (b) Triplet: $\lambda _R  = 0.33$, $\Delta
_t  = 0.33$, $\Delta _d  = 0$, $\lambda _{SO}  = 0$. (c) Singlet-Triplet Mixing:
$\lambda _R = 0.33$, $\Delta _t  = 0.33$, $\Delta _d  = -0.33$,
$\lambda _{SO} = 0.143$.
 Numbers shown in colored
areas refer to the corresponding Chern numbers. For unlabeled regions, white
(grey) areas represent systems with zero (nearly zero) gap where
Chern numbers show crossover behaviors in these regions.}
\label{fig2}
\end{figure*}

%

\begin{equation}
{\tilde H}_{\Delta _d }  = \left( {\begin{array}{*{20}c}
   0 & {\Delta _{12} } & {} & {} & {} & {}  \\
   {\Delta _{12} } & 0 & {\Delta _0 } & {} & {} & {}  \\
   {} & {\Delta _0 } & 0 & {\Delta _{12} } & {} & {}  \\
   {} & {} & {\Delta _{12} } & 0 & {\Delta _0 } & {}  \\
   {} & {} & {} & {\Delta _0 } & 0 & {\Delta _{12} }  \\
   {} & {} & {} & {} & {\Delta _{12} } & 0  \\
\end{array}} \right),
\end{equation}

\begin{equation}
{\tilde H}_{\Delta _t }  = {\tilde H}_R \left( {\lambda _R  \to \Delta _t } \right),
\end{equation}

\begin{equation}
{\tilde H}_B  = \mu _B B_z {\tilde I},
\end{equation}

\begin{equation}
{\tilde H}_\mu   = \mu {\tilde I}.
\end{equation} Blank entries in above matrices are zeros and $\tilde I$ represents unit matrix.
The matrix elements are then given by the following expressions:

\begin{equation}
\begin{array}{l}
 \alpha  = 2\cos \left( {{{\sqrt 3 k} \mathord{\left/
 {\vphantom {{\sqrt 3 k} 2}} \right.
 \kern-\nulldelimiterspace} 2}} \right){,}\\
 \beta _{\rm 1}  = 2\lambda _{SO} \sin \left( {{{\sqrt 3 k} \mathord{\left/
 {\vphantom {{\sqrt 3 k} 2}} \right.
 \kern-\nulldelimiterspace} 2}} \right),{\rm  }\beta _{\rm 2}  =  - 2\lambda _{SO} \sin \left( {\sqrt 3 k} \right){,} \\
 \gamma _1  = i\sqrt 3 \sin \left( {{{\sqrt 3 k} \mathord{\left/
 {\vphantom {{\sqrt 3 k} 2}} \right.
 \kern-\nulldelimiterspace} 2}} \right),\gamma _{\rm 2}  = i\cos \left( {{{\sqrt 3 k} \mathord{\left/
 {\vphantom {{\sqrt 3 k} 2}} \right.
 \kern-\nulldelimiterspace} 2}} \right){\rm , }\\

\gamma ' = \gamma _1  - \gamma _2 ,\gamma '' = \gamma _1  +
\gamma_2{,}\gamma  = i {,}\\

 \Delta _0  = {\Delta _d },{\rm  }\Delta _1  = { \Delta _d }e^{{i2\pi } / 3},\\{\rm  }\Delta _2  ={ \Delta _d }e^{{i4\pi } / 3} , \Delta _{12}  = \Delta _1  + \Delta _2
{.} \\
 \end{array}
\end{equation} Here $k = k_y $, and $\Delta _{i = 0,1,2} $ are the matrix elements
of $d$+$id$-wave pairing in ZGNR system \cite{Sun}. The
eigenvectors, $\Psi$, corresponding to Hamiltonian matrix Eq. (15)
have the form

\begin{equation}
 \Psi  = \left(
{\begin{array}{*{20}c}
   {\Psi _{k\bar  \uparrow } } & {\Psi _{k\bar  \downarrow } } & {\Psi _{ - k\bar  \uparrow } } & {\Psi _{ - k\bar  \downarrow } }  \\
\end{array}} \right)^T,
\end{equation} where $\Psi _{ \pm k\bar \sigma  = \bar  \uparrow ,\bar  \downarrow }  = \Psi _{ \pm k\bar \sigma  = \bar  \uparrow ,\bar  \downarrow } \left( {x;A',B'} \right)$.
Here, $x$, ${\bar \sigma }$ and $A'$($B'$) are position across the
ribbon (see figure 2), quasiparticle and sublattice pseudospin,
respectively. Note that, the quasiparticle pseudospin indices $\bar
\sigma  = \bar \sigma \left( {\bar  \uparrow ,\bar  \downarrow }
\right) $ are linear combination of original up and down spin
states.

\begin{figure}[b]%
\includegraphics*[width=0.5\textwidth]{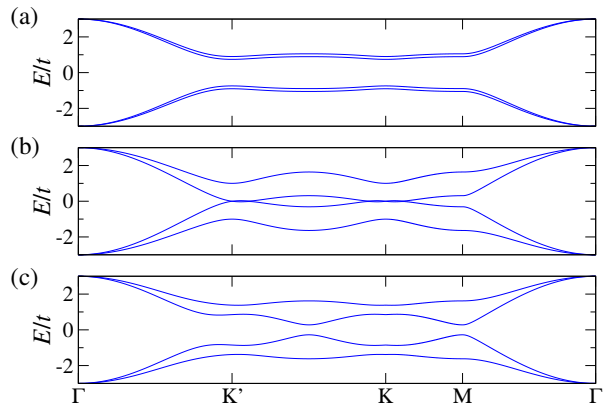}
\caption{Dispersion relation of a bulk system along different high
symmetry points (only states close to zero energy are shown). We
take $t = 1$, $\mu  =  - 3$, $\mu _B B_z  =  - 3$ for (a) singlet pairing:
$\Delta _d  =  -0.33$, $\lambda _{SO}  = 0.143$, $\lambda _R  = 0$,
$\Delta _t  = 0$. (b) triplet pairing: $\Delta _d  =  0$, $\lambda _{SO}
= 0$, $\lambda _R  = 0.33$, $\Delta _t  = 0.33$. (c) mixing: $\Delta
_d  =  -0.33$, $\lambda _{SO}  = 0.143$, $\lambda _R  = 0.33$,
$\Delta _t  = 0.33$.
}
\label{fig3}
\end{figure}

\section{Results}

\subsection{General bulk topological phase diagram}
The NCS supports MF zero energy modes at the edge of vortex cores in
the presence of a magnetic field \cite{Sato09}. To study possible
topological phases and phase transitions, we first examine the
topology of bulk band by computing Chern numbers ($C$). The gap
closing condition is used in the bulk spectrum to determine distinct
topological nature \cite{BlackSchaffer}. Topologically non-trivial
systems carry nonzero Chern numbers. Via bulk-edge correspondence,
non-trivial bulk Chern number indicates the number of edge states.
The sign of Chern number indicates the curvature of the Fermi
surface or the chirality of the bulk band gap function
\cite{Sato09}.

The topological phase diagrams as a function of the chemical
potential ($\mu$) and Zeeman coupling ($\mu _B B_z $) are presented
in figure 3. We consider three cases: (i) finite KM coupling and
spin singlet $d$+$id$-wave paring (see figure 3(a)) (ii) finite
Rashba coupling and spin triplet $p$+$ip$-wave pairing (see figure
3(b)) and (iii) mixture of above two cases (see figure 3(c)).

\begin{figure*}[t!]%
{%
\includegraphics*[width=0.33\linewidth]{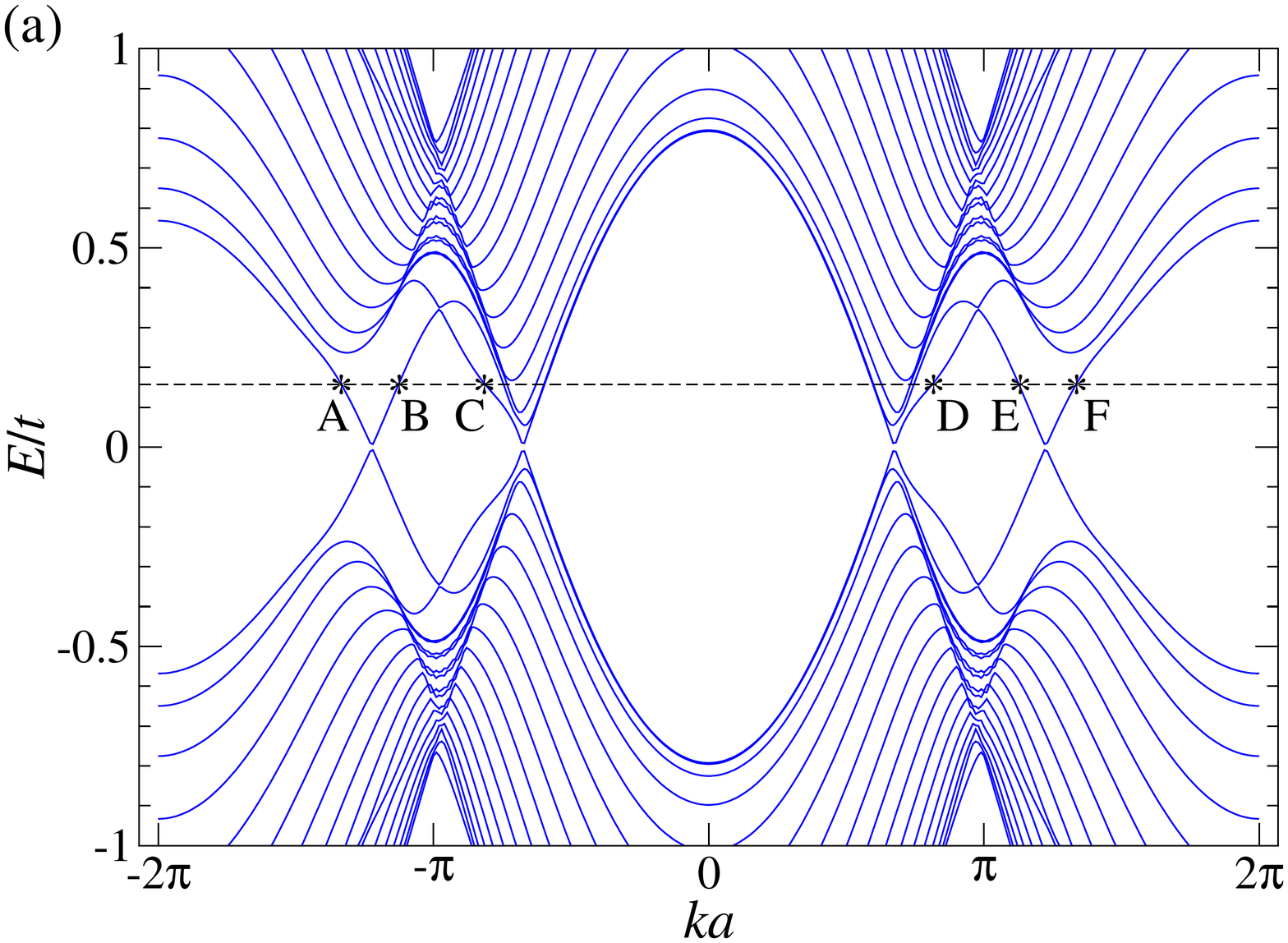}}\hfill
{%
\includegraphics*[width=0.33\linewidth]{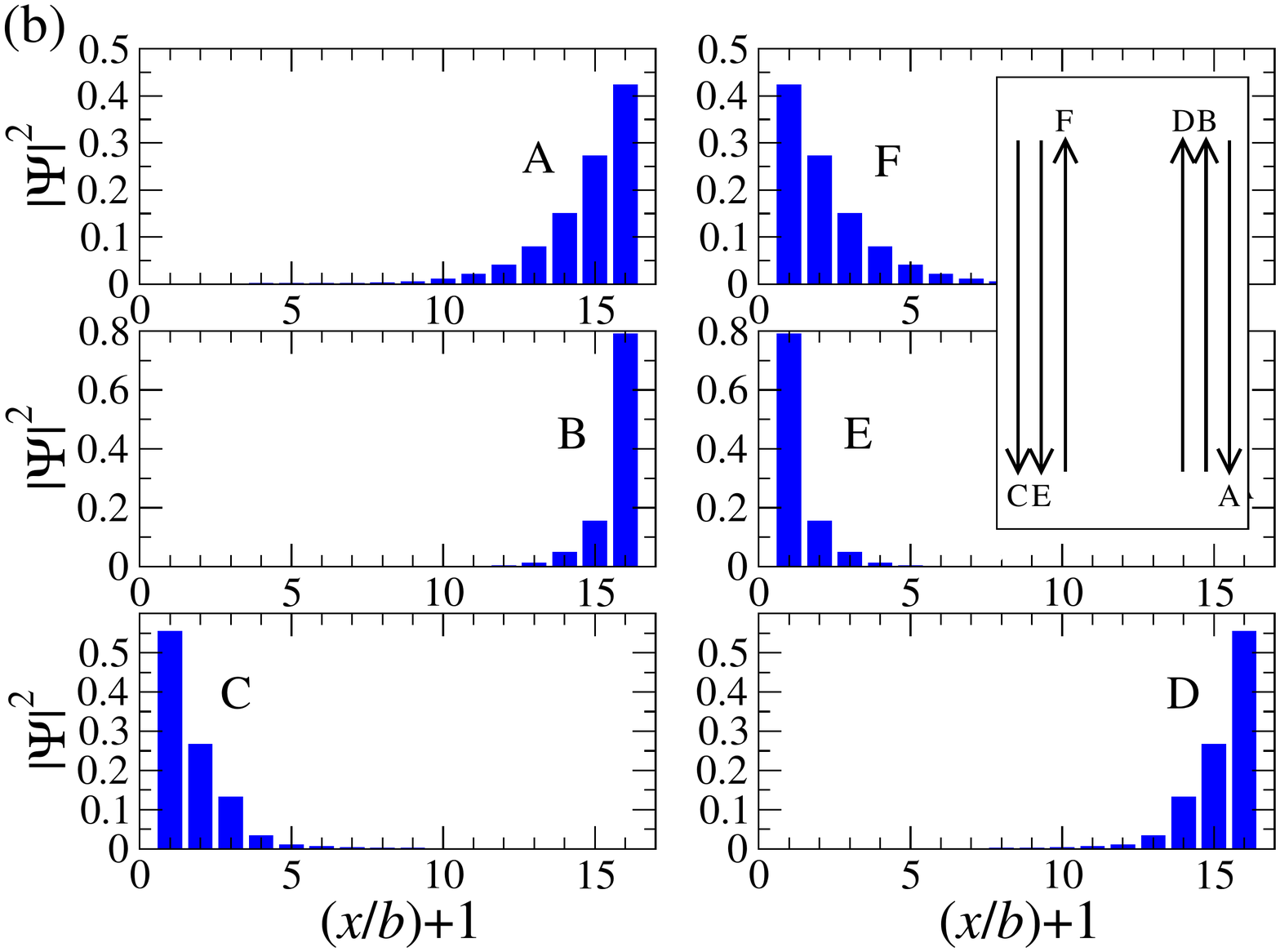}}\hfill
{\label{figs4}%
\includegraphics*[width=0.33\linewidth]{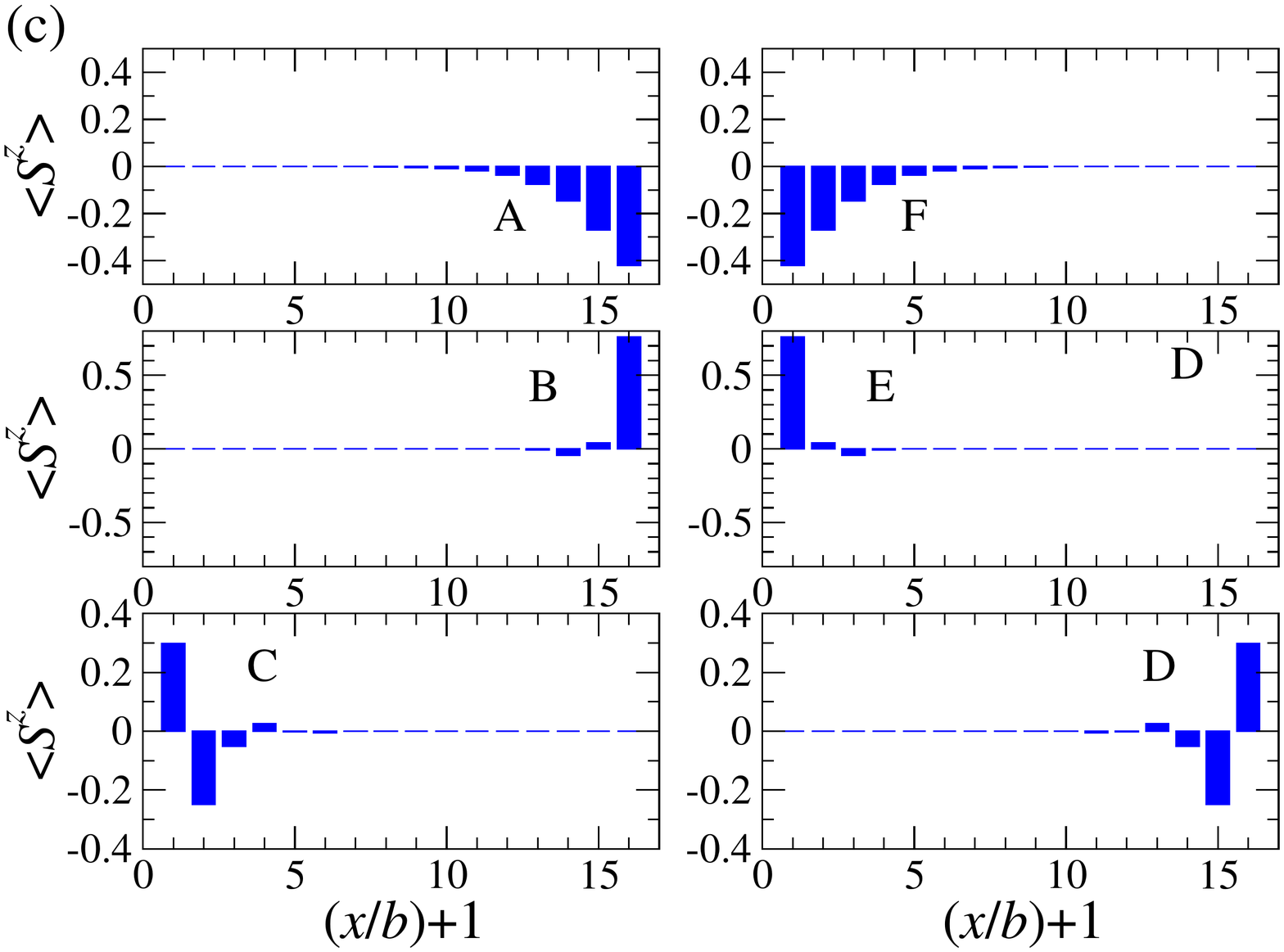}}%
\caption{(a) Band structure of a 16-ZNR for $\lambda _R  = 0.3$,
$\Delta _t  = 0.3$, $\Delta _d  =  -0.3$, $\lambda _{SO}  = 0.13$,
$\mu  =  - 3.41$, $\mu _B B_z  =  - 2.95$. The magnitude of
parameters are similar to those of bulk in figure 3(c). The Fermi
level is indicated by the dashed horizontal line above zero energy,
and the intersections with the edge state dispersion are denoted by
capital letters ("*"). Small gaps in the edge states are due to finite-size
effects \cite{BlackSchaffer}. (b) Edge state probability, charge
current (inset) distributions. The arrows in inset describe the
propagating directions of charge currents. (c) Spin polarization as a
function of position across the ZNR. The amplitude square of edge
state wave functions exhibits exponential decay from both edges into
the bulk. Here, one chiral (C, D) and one pair of helical edge
modes, (A, F) and (B, E), emerge. Due to the influence of edge mode
(C, D), B and E are not purely spin filtered edge states.}
\label{fig4}
\end{figure*}

\begin{figure}[b!]%
\includegraphics*[width=0.5\textwidth]{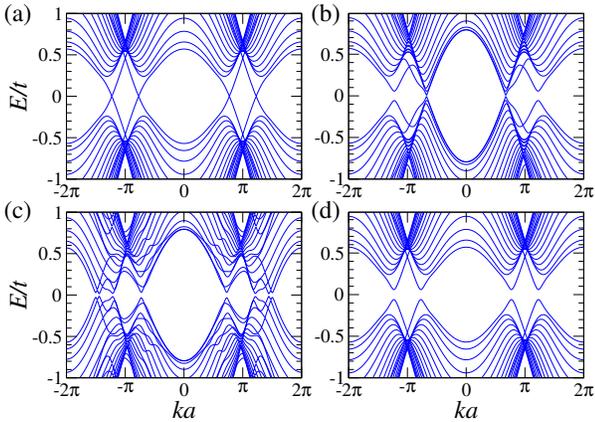}
\caption{Band structure of a 16-ZNR for different setting from figure
5(a). (a) $\lambda _R = 0$, $\Delta _t  = 0$, $\Delta _d = 0$,
$\lambda _{SO} = 0.13$, $\mu = - 3.41$, $\mu _B B_z  =  - 2.95$. (b)
$\lambda _R  = 0$, $\Delta _t = 0$, $\Delta _d  =  -0.3$, $\lambda
_{SO}  = 0.13$, $\mu =  - 3.41$, $\mu _B B_z =  - 2.95$. (c)
$\lambda _R  = 0.3$, $\Delta _t = 0.3$, $\Delta _d =  -0.3$,
$\lambda _{SO}  = 0$, $\mu = - 3.41$, $\mu _B B_z  =  - 2.95$. (d)
$\lambda _R  = 0.3$, $\Delta _t = 0.3$, $\Delta _d  = 0$, $\lambda
_{SO}  = 0.13$, $\mu = - 3.41$, $\mu _B B_z  = - 2.95$. These
figures show the band structure shown in figure 5(a) exists for a variety choices of parameters
in our model.}
\label{fig5}
\end{figure}

The phase diagram for system at a finite Zeeman coupling is shown in
figure 3(a). The gap closing points (white region) divide the phase
diagram into several parts. The Chern number is zero in the yellow
region, indicating trivial band insulators. The off-diagonal region
(blue area) shows Chern number $C = 2 $. This result is similar to
that for a chiral superconductor without Zeeman couplings
\cite{PRB2016}. The grey area indicates fluctuating Chern numbers,
which may lie near the phase boundaries. The phase diagram for bulk
system with Rashba interaction and $p$+$ip$-wave spin triplet paring
is shown in figure 3(b). Similar result has been found in the square
lattice with the same set of interactions \cite{Sato09}.

It has been shown that KM interaction opens up a gap at the Dirac
points in graphene \cite{Karyn10}. Therefore, the main effect of KM
interaction is lifting degeneracy of the bulk band. This enables us
to calculate Chern number at gap closing points. Based on the same
parameters used in figure 3(a) and (b), we consider system with all
couplings being finite in the figure 3(c).

Figure 4 shows the bulk energy bands, corresponding to the central
point ($\mu  = \mu _B B_z  =  - 3 $) in topological phase diagrams
of figure 3. In the case of KM and $d$+$id$-wave pairing (see figure
4(a)), flat and gapped bands appear around high symmetry points (see
figure 3(a)). In the case of Rashba and $p$+$ip$-wave
superconductivity (see figure 4(b)), each of the degenerate bands
splits into two, and band touching occurs at $K$ and $K'$ points
(see figure 3(b)). Figure 4(c) shows the co-existence of
singlet-triplet pairing (see figure 3(c)). A mixture of chiral and
helical edge states in the finite-size ribbon is found. We will
analyze the energy spectra of ZNRs in the following section.\\

\subsection{Band structures and edge states of ZNRs}
To confirm the emergence of edge mode in sample boundary with
non-trivial bulk Chern number, band structures of a finite sized ZNR
with periodic boundary conditions in the $y$-direction are
considered (see figure 2). In the normal state of graphene, a pair
of gapless counterpropagating helical edge modes exist when KM
interaction dominates \cite{Kane05}. On the other hand, the Rashba
coupling leads to co-propagating chiral edge modes in the ZGNR
spectrum \cite{Niu}. When KM and Rashba interaction are both
present, topological phase transitions can be induced under
staggered sublattice potential \cite{Karyn10} or exchange field
\cite{Guo}. In the superconducting states, however, the topological
non-trivial edge states are zero-energy self-conjugate MFs. The
chiral and helical MF modes can be supported by either KM or Rashba
interaction \cite{BlackSchaffer,Sun}.

\begin{figure}[!t]%
\includegraphics*[width=0.5\textwidth]{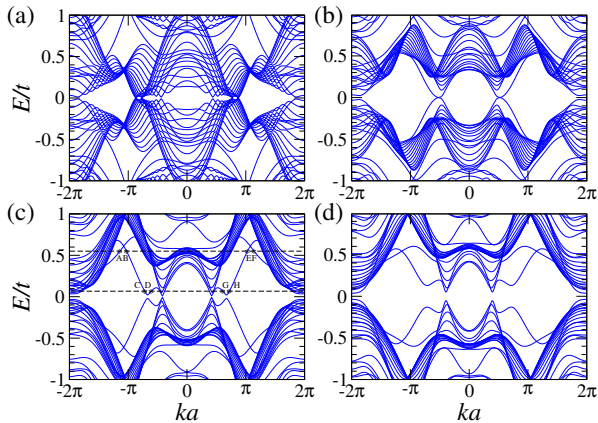}
\caption{Band structure of a 16-ZNR for $\lambda _R  = 0.3$, $\Delta
_d  =  -0.3$, $\lambda _{SO}  = 0.13$, $\mu  =  - 0.5$, $\mu _B B_z
=  - 1.5$. (a) $\Delta _t  = 0$. (b) $\Delta _t  = 0.3$. (c) $\Delta
_t  = 0.6$. (d) $\Delta _t  = 0.75$. In (c), edge states are denoted
in the same way as shown in figure 5(a). The system
undergoes a topological phase transition to a topologically non-trivial phase
as $\Delta _t $ is increased, and returns back to a topologically trivial phase at large $\Delta _t $.}
\label{fig6}
\end{figure}

The topological band structure of a 16-ZNR is shown in figure 5(a).
Figure 5(b) shows the corresponding edge state wave functions and
charge current distributions, $\left| \Psi \right|^2 = \left| {\Psi
_{{ \pm k}\bar \uparrow } } \right|^2  + \left| {\Psi _{{ \pm k}\bar
\downarrow } } \right|^2 $. Figure 5(c) shows the spin polarization
of each edge state, defined as $\left\langle {S^z } \right\rangle  =
\left| {\Psi _{{ \pm k}\bar \uparrow } } \right|^2 - \left| {\Psi
_{{ \pm k}\bar \downarrow } } \right|^2 $ \cite{NancySandler,Guo}.
It appears that there are paired (A, F), (B, E) and unpaired (C, D)
edge states in figure 5. The three edge states A, B and D are on the
same edge, and C, E and F states on the other edge (see figure
5(b)). As shown in Figure 5(c), the paired edge states (A, F) and
(B, E) belong to counter propagating helical modes. The unpaired (C,
D) edge states are co-propagating chiral modes. The oscillation of
spin polarization on unpaired edge state is similar to that found on
ZGNR with Rashba interaction \cite{NancySandler}, indicating that
this unpaired edge state comes from triplet pairing. Charge current
distributions on each edge correspond to the bulk topological phase
with $\left| C \right| = 1 $ \cite{Guo}. Note that, parameters used
here are slightly different from that in the bulk phase diagram. We
think this discrepancy is originated from the difference in
topologies of mobile $\pi$ electrons between finite and infinite
systems \cite{Nakada}. As a result, magnitude of parameters, such as
NN and NNN hopping integrals, are not the same in both cases
\cite{Hancock}.

\begin{figure}[!t]%
{%
\includegraphics*[width=0.5\textwidth]{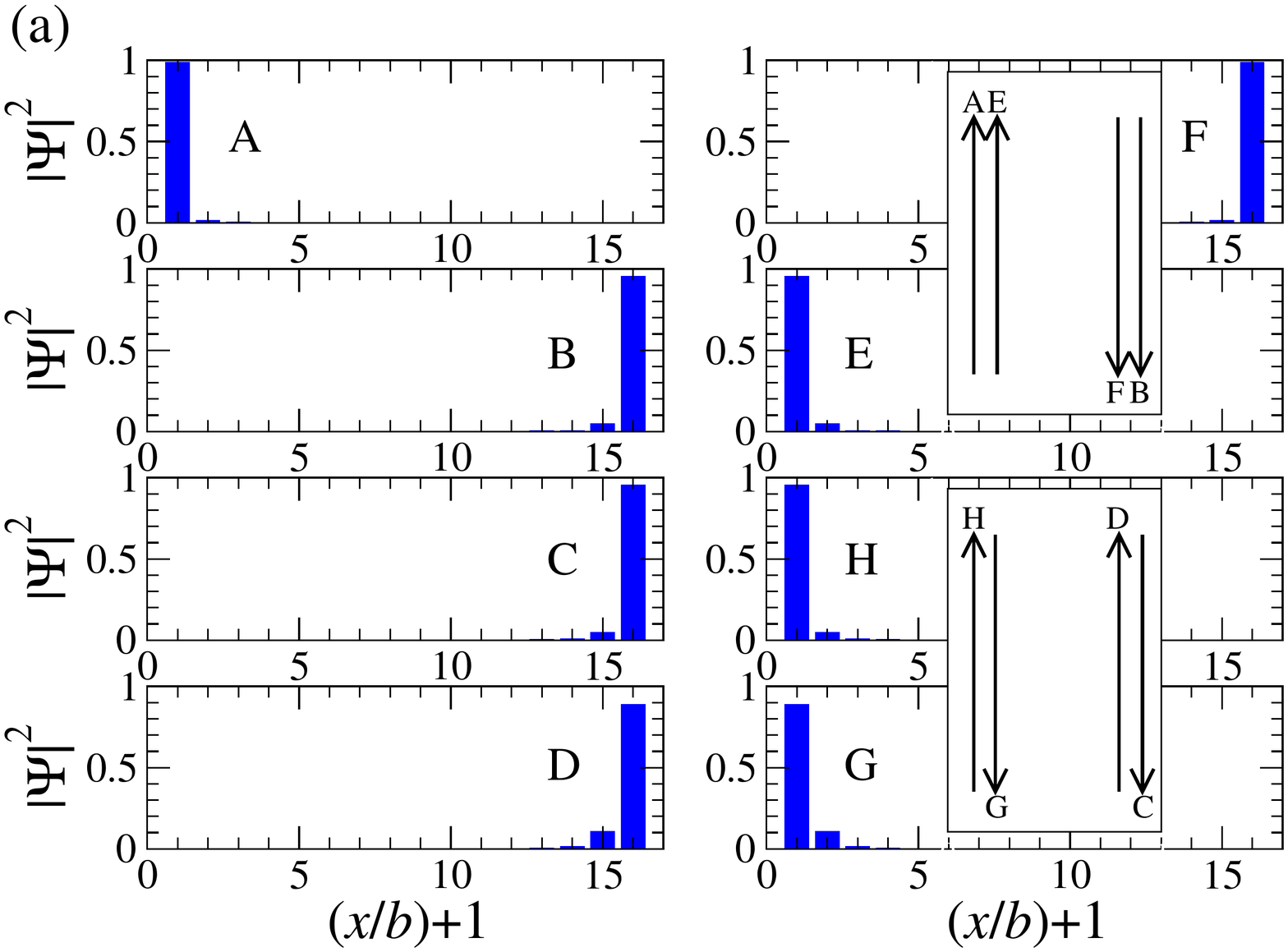}}\hfill
{\label{figs7b}%
\includegraphics*[width=0.5\textwidth]{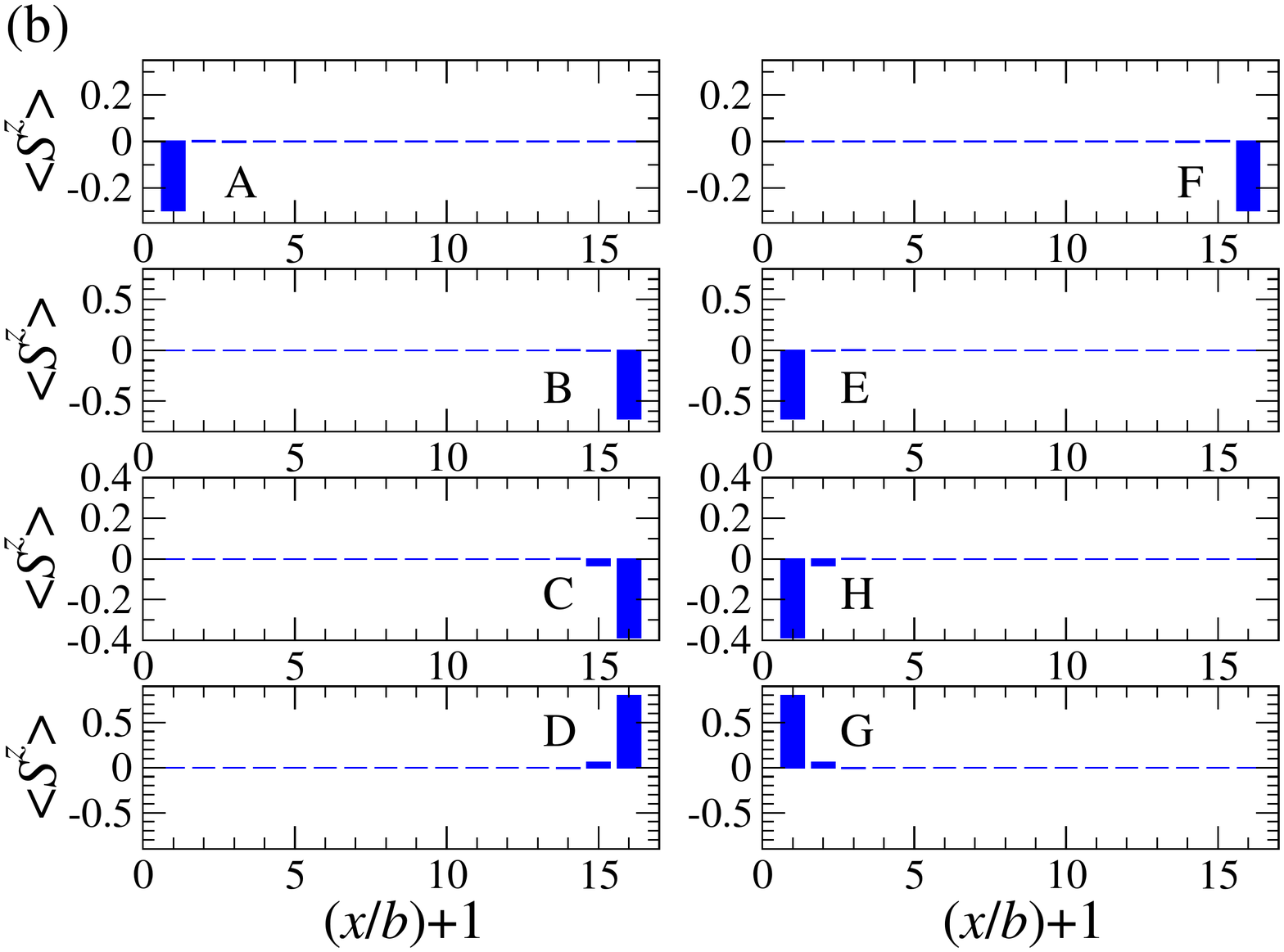}}%
\caption{(a) Schematic plot of edge state probability for chiral (upper inset), helical (lower inset) edge states
and (b) Spin polarization of figure 7(c). Edge states (A, B, E, F) and (C, D, G, H) are
associated with the intersections with the upper and lower dashed horizontal
lines in figure 7(c), respectively.}
\end{figure}

We have systematically varied the parameter setting used in figure
5(a), and the same parameters are used in figure 6. We find Dirac
cones crossing the band gap to appear in the presence of KM
interaction. The band structure shown in figure 6(a) indicates that
non-vanishing $\mu$ and $\mu _B B_z $ jointly induce TR symmetry
breaking. Each of the two Dirac cones splits into two around zone
boundary $k= \pm\pi/a$. However, we find the chiral unpaired
edge state (C, D in figure 5(a)) can also come from singlet $d$+$id$
state with TR symmetry breaking (see figure 6(b)).

To elucidate the topological phase transition for different $\Delta
_t $, we plot the corresponding band structures in figure 7. For
$\Delta _t =0$, we find no topologically edge states (see figure
7(a)). However, the system undergoes topological phase transition as
$\Delta _t $ is increased. In figure 7(b), a single chiral edge mode
appears. The situation becomes more complicated for large $\Delta
_t$ (for example, $\Delta _t  = 2\lambda _R $ in figure 7(c)). There
are two kinds of edge states appearing: upper (A, B, E, F) states at
finite energy and lower (C, D, G, H) edge states at low energy close
to zero. As shown in figure 8, the lower edge states consist of a
pair of counter propagating modes. Each edge contains spin filtered
currents. Two edge states G and H are on the same edge, and so are
the C and D states. The upper edge states belong to co-propagating
chiral edge mode at finite energies. Two edge states A and E are on
the same edge, and so are the F and B states. The spin polarization
is the same for all states, which is similar to figure 5(c).
Finally, the topological phase disappears at sufficiently large
$\Delta _t $ (see figure 7(d)).

\section{Discussions}

\subsection{Model comparisons}
It has been shown theoretically that MFs may appear in graphene. The
$d$+$id$ pairing superconducting graphene hosts two chiral edge
states in ZGNR, MFs will be created at single edge mode in the
presence of Rashba interaction and a moderate Zeeman coupling
\cite{BlackSchaffer}. Meanwhile, $d$+$id$ superconductivity in
graphene also supports helical Majorana modes in the presence of KM
interaction \cite{Sun}.

In the present work, we start by considering honeycomb lattice bulk
topological phase system with KM interaction and singlet pairing
($d$+$id$-wave) under Zeeman couplings. We further study the phase
diagram for system with Rashba interaction induced triplet pairing
($p$+$ip$-wave). It is found the outcomes in the case are similar to
that from previous studies \cite{Sato09,PRB2016}. When all
interactions are included, the phase diagrams show the feature of
singlet-triplet mixture.

We then study the edge state behavior of ZNRs in this most
interesting case. Our parity mixing model can host three states per
edge crossing the bulk gap in a ZNR. We find that helical and chiral
Majorana edge modes can exist near Fermi level simultaneously. For
large value of $\Delta_t$ ($\Delta_t = 2\lambda _R $), we also
observe chiral edge modes at finite energy, whereas helical edge
states still in the vicinity of zero energy.

\subsection{Duality}
Mathematically, there is a duality between a $s$-wave system and a
chiral $p$+$ip$-wave superconductor when Rashba SO interaction is
taken into consideration \cite{Sato09PRL}. This scheme is applied
successfully on a $s$-wave superfluid of neutral fermionic atoms in
the 2D optical square lattice with laser-field-generated effective
Rashba SO interactions \cite{Sato09PRL}. Through a unitary
transformation, the Hamiltonian with $s$-wave pairing and Rashba SO
interaction is mapped into the dual $p$+$ip$-wave Hamiltonian, which
exhibits non-trivial topological superconductivity when Zeeman
couplings go beyond a critical value \cite{Sato16}. For honeycomb
lattice with singlet $d$+$id$-wave pairing and KM interaction, it
has been demonstrated that the Hamiltonian is equivalent to a
collection of two topological ferromagnetic insulators, as described
in Section 2, offers an explanation to why such setup could exhibit
chiral edge states \cite{PRB2016}.

\subsection{Correlated honeycomb lattice}

Although many fascinating properties of graphene-based system are
well described by the low energy Dirac femions, the
electron-electron interactions are still of great interest. The
many-body effect in graphene can be induced by doping or
electron-electron interaction. The critical $M$ points in band
structure of graphene are associated with the well-known Van Hove
singularity, playing a crucial role in the pairing symmetry of
correlated graphene. For example, in graphene a spin-triplet
$f$-wave instability occurs when it is doped to the Van Hove
singularity at $1/4$ doping \cite{Honerkamp}. It hosts helical MFs
at large SO coupling \cite{Sun}, and turns into to $p$+$ip$-wave
triplet superconductor as interaction strength is increased
\cite{SCG5}. However, the issue on pairing symmetry in correlated
graphene has not been settled. Different models and computational
methods lead to various different results
\cite{SCG1,SCG2,SCG3,SCG5,SCG6,SCG4}. Therefore, it would be
interesting to further study various  possible
effects of electronic correlation on our parity mixing state.

\section{Conclusions}

In this work, we study topological phase diagram and edge states in
non-centrosymmetric superconductors on honeycomb lattice with broken
inversion symmetry. Due to the lack of inversion symmetry,
co-existence between spin-singlet and spin-triplet pairings are in
general expected in non-centrosymmetric superconductors. Promising
candidates of this kind include: graphene and graphene-based
two-dimensional materials. Due to the presence of both Kane-Mele
intrinsic and Rashba spin-orbit couplings, the superconducting state
shows parity-mixing phases between singlet $d$+$id$-wave and triplet
$p$+$ip$-wave pairing states. We compute the topological Chern
number in the bulk system and map out the topological phase diagram.
We further study the possible edge states on a finite-size ribbon.
For strong Kane-Mele intrinsic spin-orbit coupling, the helical
Majorana modes are favoured despite the underlying time-reversal
symmetry breaking $d+id$-wave singlet pairing. In the other limit
with strong Rashba spin-orbit coupling, however, the spin-triplet
$p+i p$-wave pairing is favoured, leading to chiral Majorana
fermions at edges. When the strength of the Kane-Mele and the Rashba
couplings are comparable, we find in certain parameter regime the
co-existence between chiral and helical Majorana fermions at edges,
a signature of parity-mixing. Our results provide useful guidance in
searching for non-centrosymmetric superconductors on graphene-based
materials.

\acknowledgements
 This work is supported by the MOST Grant No. 104-2112- M-009-004-MY3, the NCTS of Taiwan, R.O.C..

%
%

\end{document}